\documentclass[pra,onecolumn, showpacs, showkeys, secnumarabic, aps, amsmath, amssymb, nofootinbib, superscriptaddress, longbibliography, floatfix, table-of-contents, dblfloatfix]{revtex4-2}

\usepackage[utf8]{inputenc}
\usepackage[pdftex]{graphicx}
\usepackage{mathrsfs}
\usepackage[colorlinks, breaklinks, urlcolor={blue}, linkcolor={blue}, citecolor={blue}]{hyperref}
\usepackage{array}
\usepackage{amsmath}
\usepackage{type1cm}
\usepackage{graphicx}
\usepackage[english]{babel}
\usepackage{lmodern}
\usepackage{microtype}
\usepackage{booktabs}
\usepackage{caption}
\usepackage{braket}
\usepackage{xcolor}
\usepackage{orcidlink}

\begin{document}
\title{Teleportation of unknown qubit via  Star type tripartite states}

\author{Anushree Bhattacharjee \orcidlink{0000-0002-0655-1773} }
\email[]{a.bhattacharya.tmsl@ticollege.org}
\affiliation{Department of Mathematics, Techno Main Salt Lake (Engg. Colg.), \\Techno India Group, EM 4/1, Sector V, Salt Lake, Kolkata  700091, India}

\author{Abhijit Mandal\orcidlink{0000-0001-7101-9495}}
\email[]{a.mandal1.tmsl@ticollege.org}
\affiliation{Department of Mathematics, Techno Main Salt Lake (Engg. Colg.), \\Techno India Group, EM 4/1, Sector V, Salt Lake, Kolkata  700091, India}

\author{Sovik Roy \orcidlink{0000-0003-4334-341X} }
\email[]{s.roy2.tmsl@ticollege.org}
\affiliation{Department of Mathematics, Techno Main Salt Lake (Engg. Colg.), \\Techno India Group, EM 4/1, Sector V, Salt Lake, Kolkata  700091, India}

\begin{abstract}
\noindent Eylee Jung \textit{et.al}\cite{jung2008} had conjectured that $P_{max}=\frac{1}{2}$ is a necessary and sufficient condition for the perfect two-party teleportation and consequently the Groverian measure of entanglement for the entanglement resource must be $\frac{1}{\sqrt{2}}$. It is also known that prototype $W$ state is not useful for standard teleportation. Agrawal and Pati\cite{pati2006} have successfully executed perfect (standard) teleportation with non-prototype $W$ state. Aligned with the protocol mentioned in\cite{pati2006}, we have considered here $Star$ type tripartite states and have shown that perfect teleportation is suitable with such states. Moreover, we have taken the linear superposition of non-prototype $W$ state and its spin-flipped version and shown that it belongs to $Star$ class. Also, standard teleportation is possible with these states. It is observed that genuine tripartite entanglement is not necessary requirement for a state to be used as a channel for successful standard teleportation. We have also shown that these $Star$ class states are $P_{max}=\frac{1}{4}$ states and their Groverian entanglement is $\frac{\sqrt{3}}{2}$, thus concluding that Jung conjecture is not a necessary condition.  
\end{abstract}

\keywords{Teleportation, Groverian Entanglement, Tangle, Concurrence, Negativity, Coherence, Star State, $W$ state}

\pacs{03.65.Ud, 03.67.-a}
\maketitle
\section{Introduction:}\label{sec:introduction}
\noindent Quantum information is an emerging science and quantum teleportation is one of the many interesting protocols in this area which has garnered wide-spread prominence since its inception. The protocol was first introduced by Bennett \textit{et.al}  \textcolor{blue}{\cite{bennett1993}}.This protocol utilizes Local Operation and Classical Communication (LOCC) to transmit an unknown quantum state from sender (Alice), who possesses an unknown qubit, to a spatially separated receiver (Bob). Here, the two parties Alice and Bob share an entangled pure state (bipartite in the original paper\textcolor{blue}{\cite{bennett1993}}) or some third party may prepare such a state and distribute the qubits among Alice and Bob. The sender Alice cannot read the state otherwise the state’s quantum nature will be violated. So Alice clubs her state with the quantum channel that Alice and Bob share, consequently Alice now is in possession of two qubits and Bob retains one. To successfully execute the protocol, the Bell basis states are now considered. The joint state shared between Alice and Bob is manipulated and the two qubits of Alice (one that she desires to send and the other which comes from the shared state) in computational basis are now written in Bell basis. Then corresponding to each Bell basis states (qubits of the Bell basis are now held by Alice) there is a single qubit state (that may be the original one or a different one) in possession of Bob. At this stage, Alice performs a Bell basis measurements on her qubit and classically communicates her measurement results to Bob. On the basis of information obtained from Alice, Bob applies appropriate unitary operators to retrieve the unknown state that Alice wanted to communicate. This protocol is known as teleportation. In this protocol, the state to be teleported never exists in the channel and in the due process the state at Alice's end gets destroyed and is recreated at Bob's end. The causality is not violated in the entire protocol as classical communication is involved in the process. D. Bouwmeester \textit{et.al}\textcolor{blue}{\cite{bouwmeester1997}} later suggested how to perform experimentation on quantum teleportation  . Now a days, due to the advent of ground breaking works on quantum computer, multiparty system (or multipartite states) have come into limelight. Various teleportation schemes, henceforth were proposed with multipartite states  \textcolor{blue}{\cite{karlsson1998,man2007,Bandyopadhyay2000,Pati2002}}.
In tripartite system, there are two inequivalent classes of states, viz. $GHZ$ class and $W$ class. 
The $\vert GHZ\rangle$ and $\vert W\rangle$ states are defined as
\begin{eqnarray}
\label{ghzstate}
\vert GHZ\rangle &=& \frac{1}{\sqrt{2}}\Big(\vert 000\rangle + \vert 111\rangle\Big),
\end{eqnarray} and
\begin{eqnarray}
\label{w}
\vert W\rangle &=& \frac{1}{\sqrt{3}}\Big(\vert 001\rangle + \vert 010\rangle + \vert 100\rangle\Big).
\end{eqnarray}
Based on Stochastic Local Operations and Classical Communications (SLOCC), $\vert GHZ\rangle$ and $\vert W\rangle$ are two inequivalent classes of tripartite entangled states in the sense one cannot be converted into the other \textcolor{blue}{\cite{greenberger1,greenberger2,w1,durr2000}}. Though, $GHZ$ state (\ref{ghzstate}) can be used as channel for perfect teleportation, the $W$ state (\ref{w}) (to which we refer here as prototype $W$ state) is not useful as quantum resource for the usual teleportation protocol \textcolor{blue}{\cite{pati2006,gorvachev2003}}. Agrawal and Pati\cite{pati2006} had however shown that with non-prototypical $W$ state perfect teleportation is possible. This non-prototypical $W$ state will be discussed in the subsequent sections. In tripartite system, two other classes of states are of utmost importance. One of the states is termed as $W\tilde{W}$, which is formed by taking linear superposition of $W$ state and its spin-flipped version. The other state is known as $Star$ state which is a special type of Graph state\textcolor{blue}{\cite{buzek2003,cao2020}}. These states have been studied from various perspectives recently\textcolor{blue}{\cite{roy2023,roy2024(1),roy2024(2)}}. Yet again the states  (especially $Star$ state) are the central points of investigation in this work. Our objective of this work is to study whether the $Star$ type states can suitably be used as channel for standard teleportation. Another outlook of this paper is to study Groverian entanglement of the tripartite states considered for the standard teleportation. This motivation comes from an earlier work by Jung \textit{et.al} where they conjectured that \textit{the perfect two-party quantum teleportation is
possible if and only if the Groverian measure for the entanglement resource is  $\frac{1}{\sqrt{2}}$}\textcolor{blue}{\cite{jung2008}}. The statement tells us that if perfect two party quantum teleportation is successful via a channel then the channel has its $P_{max}$ as $\frac{1}{2}$ and Groverian entanglement as $\frac{1}{\sqrt{2}}$. The converse of this statement is also true according to Jung \textit{et.al}.\\\\ The paper is organized as follows.
In section $II$ we have discussed the perfect teleportation scheme and perform the perfect teleportation process with states such as, $GHZ$, non-prototype $W$ and its splin-flipped version, and also $Star$ types as quantum channels. In the further investigation, we have constructed a state by taking linear superposition of non-prototypical $W$ state and its spin flipped version which ultimately shown to be a member of $Star$ class. In section $III$, we have discussed the properties of the non-prototype $W$ state and that of constructed $Star$ class of states with respect to its coherence and correlations. In Section $IV$, we analyze the tangle and Groverian measure of the aforementioned class of states in section $III$, which is followed by conclusion in section $V$.
\section{Perfect teleportation protocol using tripartite states as channels:}
\subsection{Overview of perfect teleportation:}
\noindent In a teleportation protocol, Alice and Bob are positioned at different locations and share an entangled state, which serves as the quantum channel connecting them. Various types of entangled states can be utilized as this quantum channel to facilitate teleportation. The original teleportation scheme was introduced by Bennett et al. in 1993\cite{bennett1993}, and since then, numerous teleportation schemes and their applications have emerged. These schemes can be primarily categorized into two main types:
i) perfect teleportation (which we sometimes also refer to as standard teleportation in the manuscript) and ii) probabilistic teleportation schemes. In the context of perfect teleportation, the success rate of transferring the quantum state is guaranteed to be one (unity). This teleportation scheme enables the transfer of a quantum state with full fidelity- meaning the state is transferred exactly and without loss. This scheme relies on a pre-shared, maximally entangled pair of particles between the sender and receiver and deterministic steps that guarantee a successful teleportation.
The process of perfect teleportation is outlined as follows:
\begin{itemize}
\item Alice and bob share a maximally entangled pair of particles.
\item Alice measures her entangled particle along with the particle to be teleported.
\item Alice sends the outcomes of her measurements to Bob through a classical channel.
\item Bob applies a specific unitary transformation on his particle to recreate the original state.
\end{itemize}
\subsection*{GHZ state:}
\noindent The GHZ state is already defined in eq.(\ref{ghzstate}). Here we consider that the qubits are distributed among two parties Alice and Bob, while Alice retaining first two qubits and Bob takes third qubit. Suppose Alice has an unknown single qubit 
\begin{eqnarray}
\label{singlequbit}
\vert \psi\rangle_{A} = \alpha\vert 0\rangle_{A} + \beta\vert 1\rangle_{A}, \:\:|\alpha|^2 + |\beta^2| = 1.
\end{eqnarray}
Alice wants to send this qubit to Bob. The product state thus obtained is given as
\begin{eqnarray}
\label{proghz}
\vert \psi\rangle_{A}\otimes \vert GHZ\rangle_{AAB} = \frac{1}{\sqrt{2}}\Big(\alpha \vert 000\rangle_{AAA}\vert 0\rangle_{B} + \alpha\vert 011\rangle_{AAA}\vert 1\rangle_{B} + \beta \vert 100\rangle_{AAA}\vert 0\rangle_{B} + \beta \vert 111\rangle_{AAA}\vert 1\rangle_{B}\Big).
\end{eqnarray}
Next we consider the three qubit basis states as 
\begin{eqnarray}
\label{basisghz}
\vert ghz_{1}\rangle^{\pm}_{AAB} &=& \frac{1}{\sqrt{2}}\Big(\vert 000\rangle \pm \vert 111\rangle\Big)\nonumber\\
\vert ghz_{2}\rangle^{\pm}_{AAB} &=& \frac{1}{\sqrt{2}}\Big(\vert 100\rangle \pm \vert 011\rangle\Big).
\end{eqnarray}
Using eqs.(\ref{basisghz}) in (\ref{proghz}) we get
\begin{eqnarray}
\label{proghz1}
\vert \psi\rangle_{A}\otimes \vert GHZ\rangle_{AAB} = \frac{1}{2}\Big[\vert ghz_{1}\rangle^{+}_{AAA}(\alpha\vert 0\rangle_{B} + \beta\vert 1\rangle_{B}) + \vert ghz_{1}\rangle^{-}_{AAA}(\alpha\vert 0\rangle_{B} - \beta\vert 1\rangle_{B}) +\nonumber\\  \vert ghz_{2}\rangle^{+}_{AAA}(\alpha\vert 1\rangle_{B} + \beta\vert 0\rangle_{B}) + \vert ghz_{2}\rangle^{-}_{AAA}(\beta\vert 0\rangle_{B} - \alpha\vert 1\rangle_{B})\Big].
\end{eqnarray}
The table $I$ gives the measurements by Alice and the results that she needs to communicate to Bob and subsequently the unitary operators Bob applies that will retrieve the single qubit that Alice wants to send him. 
\begin{table}[h!]
\begin{center}
\caption{Summary of perfect teleportation using $GHZ$ state}
\label{table1}
\begin{tabular}{|c|c|c|}
\hline
Alice's measurement outcome & What Bob gets & What Bob needs to apply\\
\hline
$\vert ghz_{1}\rangle^{+}$ & $\alpha\vert 0\rangle_{B} + \beta\vert 1\rangle_{B}$ & $I$ gate (i.e. nothing to do)\\
\hline
$\vert ghz_{1}\rangle^{-}$ & $\alpha\vert 0\rangle_{B} - \beta\vert 1\rangle_{B}$ & $Z$ gate\\
\hline
$\vert ghz_{2}\rangle^{+}$ & $\alpha\vert 1\rangle_{B} + \beta\vert 0\rangle_{B}$ & $X$ gate\\
\hline
$\vert ghz_{2}\rangle^{-}$ & $\beta\vert 0\rangle_{B} - \alpha\vert 1\rangle_{B}$ & $iY$ gate\\
\hline
\end{tabular}
\end{center}
\end{table} 
\subsection*{Non-prototype $W$-state:}
\noindent It is known that the prototype $W$ state may not be suitable for perfect teleportation\cite{pati2006}. The class of states $\vert W_{n}\rangle_{123}$ belongs to the category of $W$ states which can be used as an entanglement resource. This class is defined as 
\begin{eqnarray}
\label{genw1}
\vert W_{n}\rangle = \frac{1}{\sqrt{2+2n}}\Big[\vert 100\rangle + \sqrt{n}e^{i\gamma}\vert 010\rangle + \sqrt{n+1}e^{i\delta}\vert 001\rangle\Big],
\end{eqnarray}
where $n$ is real number and $\gamma$ and $\delta$ are phases. By taking $n=1$ and by setting phases to zero we get,
\begin{eqnarray}
\label{genw2}
\vert W_{1}\rangle =  \frac{1}{2}\Big[\vert 100\rangle + \vert 010\rangle + \sqrt{2}\vert 001\rangle\Big]
\end{eqnarray}
We call such a state (\ref{genw2}) as non-prototypical $W$ state. We shall now show that the state (\ref{genw2}) can be used for perfect teleportation. We assume that the qubits in the state (\ref{genw2}) are shared among two parties Alice and Bob. Alice retains first two qubits and Bob takes the third qubit. Now if Alice wants to send her unknown qubit represented in eq.(\ref{singlequbit}) to Bob, she clubs her state to the channel represented by $\vert W_{1}\rangle$ so that the following is obtained.
\begin{eqnarray}
\label{prow1}
\vert \psi\rangle_{A}\otimes \vert W_{1}\rangle_{AAB} = \frac{1}{2}\Big[\alpha\vert 010\rangle_{AAA}\vert 0\rangle_{B} + \alpha\vert 001\rangle_{AAA}\vert 0\rangle_{B} + \sqrt{2}\alpha\vert 000\rangle_{AAA}\vert 1\rangle_{B} + \nonumber\\
\beta \vert 110\rangle_{AAA}\vert 0\rangle_{B} + \beta \vert 101\rangle_{AAA}\vert 0\rangle_{B} + \sqrt{2}\beta\vert 100\rangle_{AAA}\vert 1\rangle_{B}\Big].
\end{eqnarray}
Considering the basis states,
\begin{eqnarray}
\label{basisw}
\vert w_{1}\rangle^{\pm}_{AAB} &=& \frac{1}{2}\Big(\vert 010\rangle + \vert 001\rangle \pm \sqrt{2}\vert 100\rangle\Big)\nonumber\\
\vert w_{2}\rangle^{\pm}_{AAB} &=& \frac{1}{2}\Big(\vert 101\rangle + \vert 110\rangle \pm \sqrt{2}\vert 000\rangle\Big),
\end{eqnarray}
we can rewrite
\begin{eqnarray}
\label{basisw1}
\vert w_{1}\rangle^{+} + \vert w_{1}\rangle^{-} &=& \vert 010\rangle + \vert 001\rangle\nonumber\\
\vert w_{1}\rangle^{+} - \vert w_{1}\rangle^{-} &=& \sqrt{2}\vert 100\rangle \nonumber\\
\vert w_{2}\rangle^{+} + \vert w_{2}\rangle^{-} &=& \vert 110\rangle + \vert 101\rangle\nonumber\\
\vert w_{2}\rangle^{+} - \vert w_{2}\rangle^{-} &=& \sqrt{2}\vert 000\rangle.
\end{eqnarray}
Using eq.(\ref{basisw1}) in eq.(\ref{prow1}) we get
\begin{eqnarray}
\label{prow11}
\vert \psi\rangle_{A}\otimes \vert W_{1}\rangle_{AAB} = \frac{1}{2}\Big[\vert w_{1}\rangle^{+}_{AAA}(\alpha\vert 0\rangle_{B} + \beta \vert 1\rangle_{B}) + \vert w_{1}\rangle^{-}_{AAA}(\alpha\vert 0\rangle_{B} - \beta \vert 1\rangle_{B}) +\nonumber\\
\vert w_{2}\rangle^{+}_{AAA}(\alpha\vert 1\rangle_{B} + \beta \vert 0\rangle_{B}) + \vert w_{2}\rangle^{-}_{AAA}(\beta\vert 0\rangle_{B} - \alpha \vert 1\rangle_{B})\Big].
\end{eqnarray}
As before, the table $II$ shows the measurements by Alice and her communication to Bob and subsequently the unitary gates Bob applies that will retrieve the single qubit that Alice wants to send him. 
\begin{table}[h!]
\begin{center}
\caption{Summary of perfect teleportation using $W_{1}$ state}
\label{table2}
\begin{tabular}{|c|c|c|}
\hline
Alice's measurement outcome & What Bob gets & What Bob needs to apply\\
\hline
$\vert w_{1}\rangle^{+}$ & $\alpha\vert 0\rangle_{B} + \beta\vert 1\rangle_{B}$ & $I$ gate (i.e. nothing to do)\\
\hline
$\vert w_{1}\rangle^{-}$ & $\alpha\vert 0\rangle_{B} - \beta\vert 1\rangle_{B}$ & $Z$ gate\\
\hline
$\vert w_{2}\rangle^{+}$ & $\alpha\vert 1\rangle_{B} + \beta\vert 0\rangle_{B}$ & $X$ gate\\
\hline
$\vert w_{2}\rangle^{-}$ & $\beta\vert 0\rangle_{B} - \alpha\vert 1\rangle_{B}$ & $iY$ gate\\
\hline
\end{tabular}
\end{center}
\end{table}
\subsection*{$\tilde{W_{1}}$ state:}
We now consider the spin-flipped version of the $W_{1}$ state of eq.(\ref{genw2}), which is given by
\begin{eqnarray}
\label{genw3}
\vert \tilde{W_{1}}\rangle =  \frac{1}{2}\Big[\vert 011\rangle + \vert 101\rangle + \sqrt{2}\vert 110\rangle\Big].
\end{eqnarray}
As before the first two qubits of the state (\ref{genw3}) belongs to Alice and the third qubit belongs to Bob. Moreover, Alice wants to send the unknown qubit $\vert \psi\rangle$ of eq.(\ref{singlequbit}), that is in her possession, to Bob. After clubbing her unknown qubit, the following is obtained.
\begin{eqnarray}
\label{prow12}
\vert \psi\rangle_{A}\otimes \vert \tilde{W_{1}}\rangle_{AAB} = \frac{1}{2}\Big[\alpha\vert 001\rangle_{AAA}\vert 1\rangle_{B} + \alpha\vert 010\rangle_{AAA}\vert 1\rangle_{B} + \sqrt{2}\alpha\vert 011\rangle_{AAA}\vert 0\rangle_{B} + \nonumber\\
\beta \vert 101\rangle_{AAA}\vert 1\rangle_{B} + \beta \vert 110\rangle_{AAA}\vert 1\rangle_{B} + \sqrt{2}\beta\vert 111\rangle_{AAA}\vert 0\rangle_{B}\Big].
\end{eqnarray}
Let us now take the basis states as
\begin{eqnarray}
\label{basiswbar}
\vert \tilde{w_{1}}\rangle^{\pm}_{AAB} &=& \frac{1}{2}\Big(\vert 001\rangle + \vert 010\rangle \pm \sqrt{2}\vert 111\rangle\Big)\nonumber\\
\vert \tilde{w_{2}}\rangle^{\pm}_{AAB} &=& \frac{1}{2}\Big(\vert 101\rangle + \vert 110\rangle \pm \sqrt{2}\vert 011\rangle\Big).
\end{eqnarray}
Rearranging the terms in eq.(\ref{basiswbar}) we have
\begin{eqnarray}
\label{basiswbar1}
\vert \tilde{w_{1}}\rangle^{+} + \vert \tilde{w_{1}}\rangle^{-} &=& \vert 001\rangle + \vert 010\rangle\nonumber\\
\vert \tilde{w_{1}}\rangle^{+} - \vert \tilde{w_{1}}\rangle^{-} &=& \sqrt{2}\vert 111\rangle \nonumber\\
\vert \tilde{w_{2}}\rangle^{+} + \vert \tilde{w_{2}}\rangle^{-} &=& \vert 101\rangle + \vert 110\rangle\nonumber\\
\vert \tilde{w_{2}}\rangle^{+} - \vert \tilde{w_{2}}\rangle^{-} &=& \sqrt{2}\vert 011\rangle.
\end{eqnarray}
Using eq.(\ref{basiswbar1}) in eq.(\ref{prow12}) we get
\begin{eqnarray}
\label{prow111}
\vert \psi\rangle_{A}\otimes \vert \tilde{W_{1}}\rangle_{AAB} = \frac{1}{2}\Big[\vert \tilde{w_{1}}\rangle^{+}_{AAA}(\alpha\vert 1\rangle_{B} + \beta \vert 0\rangle_{B}) + \vert \tilde{w_{1}}\rangle^{-}_{AAA}(\alpha\vert 1\rangle_{B} - \beta \vert 0\rangle_{B}) +\nonumber\\
\vert \tilde{w_{2}}\rangle^{+}_{AAA}(\alpha\vert 0\rangle_{B} + \beta \vert 1\rangle_{B}) + \vert \tilde{w_{2}}\rangle^{-}_{AAA}(-\alpha\vert 0\rangle_{B} + \beta \vert 1\rangle_{B})\Big].
\end{eqnarray}
For successful running of the perfect teleportation Alice and Bob must adhere to the settings as depicted in table $III$.
\begin{table}[h!]
\begin{center}
\caption{Summary of perfect teleportation using $\tilde{W}$ state}
\label{table3}
\begin{tabular}{|c|c|c|}
\hline
Alice's measurement outcome & What Bob gets & What Bob needs to apply\\
\hline
$\vert \tilde{w_{1}}\rangle^{+}$ & $\alpha\vert 1\rangle_{B} + \beta\vert 0\rangle_{B}$ & $X$ gate\\
\hline
$\vert \tilde{w_{1}}\rangle^{-}$ & $\alpha\vert 1\rangle_{B} - \beta\vert 0\rangle_{B}$ & $-iY$ gate\\
\hline
$\vert \tilde{w_{2}}\rangle^{+}$ & $\alpha\vert 0\rangle_{B} + \beta\vert 1\rangle_{B}$ & $I$ gate (i.e. nothing to do)\\
\hline
$\vert \tilde{w_{2}}\rangle^{-}$ & $-\alpha\vert 0\rangle_{B} + \beta \vert 1\rangle_{B}$ & $-Z$ gate\\
\hline
\end{tabular}
\end{center}
\end{table}\newpage
\noindent Before we proceed further, we shall analyze two different tripartite classes of states which have significant importance in information processing. These two class of states are respectively $W\tilde{W}$ class and $Star$ class of states, which are defined as 
\begin{eqnarray}
\label{wwbar}
\vert W\tilde{W}\rangle = \frac{1}{\sqrt{2}}\Big(\vert W\rangle + \vert \tilde{W}\rangle\Big),
\end{eqnarray} and 
\begin{eqnarray}
\label{star}
\vert Star\rangle_{ABC} = \frac{1}{2}\Big(\vert 000\rangle_{ABC} + \vert 100\rangle_{ABC} + \vert 101\rangle_{ABC} + \vert 111\rangle_{ABC}\Big).
\end{eqnarray}
In eq.(\ref{wwbar}), $\vert \tilde{W}\rangle = \frac{1}{\sqrt{3}}\Big(\vert 110\rangle + \vert 101\rangle + \vert 011\rangle\Big)$ is the spin-flipped version of $W$ state of eq. (\ref{w}). It is seen that $\vert W\tilde{W}\rangle$ is the equal linear superposition of a standard $W$ state, which is $\frac{1}{\sqrt{3}}\Big[\vert 001\rangle + \vert 010\rangle + \vert 100\rangle\Big]$ and its spin flipped version. This tripartite state has both bipartite and tripartite distribution of entanglement. On the other hand, there is $Star$ state of eq.(\ref{star}), which is not symmetric, since the correlations of these states are present in an asymmetric way. There are two types of qubits in $Star$ state viz. (i) two peripheral qubits and (ii) one central qubit. The first two qubits ($A$ and $B$) of the state are peripheral and the third qubit ($C$) is the central qubit. This means that when central qubit is traced out, the remaining qubits are left in a separable state whereas if we take partial trace over first and second qubit (i.e. with respect to peripheral qubits), entanglement is still present in the remaining qubits. In view of this characteristic, Star state can be distinguished from other types of three qubit states. A pictorial representation of Star state showing peripheral qubits and central qubit are shown in the following Fig $I$.
\begin{figure}[hbtp]
\centering
\resizebox{5cm}{3cm}{\includegraphics{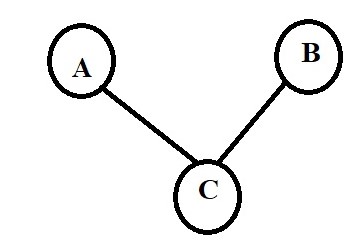}}
\caption{The pictorial representation of Star class of states showing peripheral qubits $A$ and $B$ and the central qubit$C$.}
\end{figure}\\
Let us now discuss about how these two classes of states contribute in the information processing such as teleportation. It can be easily shown that just like usual $W$ state is not useful for teleportation and so is $W\tilde{W}$. Likewise the Star state given in eq.(\ref{star}) is also not useful for standard teleportation. Hence we take another form of Star state with a phase difference and show that with this state perfect teleportation can be successfully conducted.  
\subsection*{$Star^1$ state:} 
\noindent The Star state for our study is considered to be
\begin{eqnarray}
\label{star1}
\vert Star^1\rangle_{AAB} = \frac{1}{2}\Big(\vert 000\rangle_{AAB} + \vert 100\rangle_{AAB} + \vert 101\rangle_{AAB} - \vert 111\rangle_{AAB}\Big)
\end{eqnarray}
Alice takes the possession of first two peripheral qubits and Bob takes the third qubit which is central qubit. Alice wants to send her unknown state (which is defined in eq.(\ref{singlequbit}) ) to Bob. After clubbing this state with the channel (\ref{star1}), we get the following.
\begin{eqnarray}
\label{star11}
\vert \psi\rangle_{A}\otimes \vert Star^1\rangle_{AAB} = \frac{1}{2}\Big[\alpha\vert 000\rangle_{AAA}\vert 0\rangle_{B} + \alpha\vert 010\rangle_{AAA}\vert 0\rangle_{B} + \alpha\vert 010\rangle_{AAA}\vert 1\rangle_{B} -\alpha\vert 011\rangle_{AAA}\vert 1\rangle_{B}\nonumber\\
\beta\vert 100\rangle_{AAA}\vert 0\rangle_{B} + \beta\vert 110\rangle_{AAA}\vert 0\rangle_{B} + \beta\vert 110\rangle_{AAA}\vert 1\rangle_{B} -\beta\vert 111\rangle_{AAA}\vert 1\rangle_{B}\Big].
\end{eqnarray}
Let us consider the following four basis states.
\begin{eqnarray}
\label{starbasis}
\vert s_{1}\rangle^{+} &=& \frac{1}{2}\Big[\vert 000\rangle + \vert 010\rangle + \vert 110\rangle - \vert 111\rangle\Big]\nonumber\\
\vert s_{1}\rangle^{-} &=& \frac{1}{2}\Big[\vert 000\rangle + \vert 010\rangle - \vert 110\rangle + \vert 111\rangle\Big]\nonumber\\
\vert s_{2}\rangle^{+} &=& \frac{1}{2}\Big[\vert 100\rangle + \vert 110\rangle + \vert 010\rangle - \vert 011\rangle\Big]\nonumber\\
\vert s_{2}\rangle^{-} &=& \frac{1}{2}\Big[\vert 100\rangle + \vert 110\rangle - \vert 010\rangle + \vert 011\rangle\Big],
\end{eqnarray}
so that we have,
\begin{eqnarray}
\label{starbasis2}
\vert s_{1}\rangle^{+} + \vert s_{1}\rangle^{-} &=& \vert 000\rangle + \vert 010\rangle\nonumber\\
\vert s_{1}\rangle^{+} - \vert s_{1}\rangle^{-} &=& \vert 110\rangle - \vert 111\rangle\nonumber\\
\vert s_{2}\rangle^{+} + \vert s_{2}\rangle^{-} &=& \vert 100\rangle + \vert 110\rangle\nonumber\\
\vert s_{2}\rangle^{+} - \vert s_{2}\rangle^{-} &=& \vert 010\rangle - \vert 011\rangle.
\end{eqnarray}
Using eq.(\ref{starbasis2}) in (\ref{star11}) we get
\begin{eqnarray}
\label{starr}
\vert \psi\rangle_{A}\otimes \vert Star^1\rangle_{AAB} = \frac{1}{2}\Big[\vert s_{1}\rangle^{+}_{AAA} (\alpha\vert 0\rangle_{B} + \beta\vert 1\rangle_{B}) + \vert s_{1}\rangle^{-}_{AAA}(\alpha\vert 0\rangle_{B} - \beta\vert 1\rangle_{B}) \nonumber\\+ \vert s_{2}\rangle^{+}_{AAA}(\alpha\vert 1\rangle_{B} + \beta \vert 0\rangle_{B}) + \vert s_{2}\rangle^{-}_{AAA}(-\alpha \vert 1\rangle_{B} + \beta \vert 0\rangle_{B})\Big].
\end{eqnarray}
For successful running of the perfect teleportation Alice and Bob must adhere to the following settings shown in table $IV$.
\begin{table}[h!]
\begin{center}
\caption{Summary of perfect teleportation using $Star^{1}$ state}
\label{table4}
\begin{tabular}{|c|c|c|}
\hline
Alice's measurement outcome & What Bob gets & What Bob needs to apply\\
\hline
$\vert s_{1}\rangle^{+}$ & $\alpha\vert 0\rangle_{B} + \beta\vert 1\rangle_{B}$ & $I$ gate (i.e. nothing to do)\\
\hline
$\vert s_{1}\rangle^{-}$ & $\alpha\vert 0\rangle_{B} - \beta\vert 1\rangle_{B}$ & $Z$ gate\\
\hline
$\vert s_{2}\rangle^{+}$ & $\alpha\vert 1\rangle_{B} + \beta\vert 0\rangle_{B}$ & $X$ gate \\
\hline
$\vert s_{2}\rangle^{-}$ & $-\alpha\vert 1\rangle_{B} + \beta \vert 0\rangle_{B}$ & $iY$ gate\\
\hline
\end{tabular}
\end{center}
\end{table}
\subsection*{$S_{w\tilde{w}}$ state:}
\noindent We shall now construct a state by taking linear superposition of $W_{1}$ state (which we have already referred as non-prototypical $W$ state in eq.(\ref{genw2})) and its spin-flipped version $\tilde{W_{1}}$. Let us denote this state by $S_{w\tilde{w}}$ and is represented as
\begin{eqnarray}
\label{starwwbar}
\vert S_{w\tilde{w}}\rangle_{AAB} = \frac{1}{\sqrt{2}}\Big(\vert W_{1}\rangle_{AAB} + \vert \tilde{W_{1}}\rangle_{AAB}\Big).
\end{eqnarray}
The given state belongs to $Star$ class\footnote{The analysis has been shown in subsequent section.}. The state (\ref{starwwbar}) has coherence and correlation distributed at all possible levels. Moreover, just as in the case of $Star$ states of eq. (\ref{star}), the state $S_{w\tilde{w}}$ has one central qubit (retained by Bob the receiver) and two peripheral qubits (hold by Alice the sender). Hence the state presented in eq.(\ref{starwwbar}) belongs to $Star$ class. \\\\

\noindent We will now show that the state $S_{w\tilde{w}}$ is also useful for perfect teleportation.
Let us suppose that Alice and Bob share the state $S_{w\tilde{w}}$ where first two qubits are retained by Alice and the third qubit is kept by Bob. The explicit representation of such a state is shown below.
\begin{eqnarray}
\label{starwwbar1}
\vert S_{w\tilde{w}}\rangle_{AAB} = \frac{1}{2\sqrt{2}}\Big(\vert 100\rangle_{AAB} + \vert 010\rangle_{AAB} + \sqrt{2}\vert 001\rangle_{AAB} + \vert 011\rangle_{AAB} + \vert 101\rangle_{AAB} + \sqrt{2}\vert 110\rangle_{AAB} \Big).
\end{eqnarray}
Alice wants to send single unknown qubit $\vert \psi\rangle$ of eq.(\ref{singlequbit}) to Bob as before. So when this qubit is clubbed with the channel (\ref{starwwbar1}), the resulting state is
\begin{eqnarray}
\label{prostarwwbar}
\vert \psi\rangle_{A}\otimes \vert  S_{w\tilde{w}}\rangle_{AAB} = \frac{1}{2\sqrt{2}}\Big[\alpha(\vert 010\rangle + \vert 001\rangle)_{AAA}\vert 0\rangle_{B} + \alpha \sqrt{2}\vert 000\rangle_{AAA}\vert 1\rangle_{B} + \alpha(\vert 001\rangle + \vert 010\rangle)_{AAA}\vert 1\rangle_{B}+\nonumber\\ \alpha \sqrt{2}\vert 011\rangle\vert 0\rangle_{B} + \beta(\vert 110\rangle + \vert 101\rangle)_{AAA}\vert 0\rangle_{B} + \beta\sqrt{2}\vert 100\rangle\vert 1\rangle_{B} +\beta(\vert 101\rangle + \vert 110\rangle)_{AAA}\vert 1\rangle_{B} + \beta \sqrt{2}\vert 111\rangle_{AAA}\vert 0\rangle_{B}\Big]\nonumber\\.
\end{eqnarray}
Using eqs.(\ref{basisw}) and (\ref{basiswbar}) as basis elements, the eq.(\ref{prostarwwbar}) can be rewritten as,
\begin{eqnarray}
\label{prostarwwbar1}
\vert \psi\rangle_{A}\otimes \vert  S_{w\tilde{w}}\rangle_{AAB} = \frac{1}{2\sqrt{2}}\Big[\vert w_{1}\rangle^{+}_{AAA}(\alpha\vert 0\rangle_{B} + \beta\vert 1\rangle_{B}) + \vert w_{1}\rangle^{-}_{AAA}(\alpha\vert 0\rangle_{B} - \beta\vert 1\rangle_{B}) +\nonumber\\ \vert w_{2}\rangle^{+}_{AAA}(\alpha\vert 1\rangle_{B} + \beta\vert 0\rangle_{B}) + \vert w_{2}\rangle^{-}_{AAA}(-\alpha\vert 1\rangle_{B} + \beta\vert 0\rangle_{B}) +\nonumber\\ \vert \tilde{w_{1}}\rangle^{+}_{AAA}(\alpha\vert 1\rangle_{B}+\beta\vert 0\rangle_{B}) + \vert \tilde{w_{1}}\rangle^{-}_{AAA}(\alpha\vert 1\rangle_{B}-\beta\vert 0\rangle_{B})+\nonumber\\ \vert \tilde{w_{2}}\rangle^{+}_{AAA}(\alpha\vert 0\rangle_{B}+\beta\vert 1\rangle_{B}) + \vert \tilde{w_{2}}\rangle^{-}_{AAA}(-\alpha\vert 0\rangle_{B}+\beta\vert 1\rangle_{B})\Big].
\end{eqnarray}
For successful perfect teleportation, Alice and Bob will go for the following settings, as shown in table $V$.
\begin{table}[h!]
\begin{center}
\caption{Summary of perfect teleportation using $S_{w\tilde{w}}$ state}
\label{table5}
\begin{tabular}{|c|c|c|}
\hline
Alice's measurement outcome & What Bob gets & What Bob needs to apply\\
\hline
$\vert w_{1}\rangle^{+}$ & $\alpha\vert 0\rangle_{B} + \beta\vert 1\rangle_{B}$ & $I$ gate (i.e. nothing to do)\\
\hline
$\vert w_{1}\rangle^{-}$ & $\alpha\vert 0\rangle_{B} - \beta\vert 1\rangle_{B}$ & $Z$ gate\\
\hline
$\vert w_{2}\rangle^{+}$ & $\alpha\vert 1\rangle_{B} + \beta\vert 0\rangle_{B}$ & $X$ gate \\
\hline
$\vert w_{2}\rangle^{-}$ & $-\alpha\vert 1\rangle_{B} + \beta \vert 0\rangle_{B}$ & $iY$ gate\\
\hline
$\vert \tilde{w_{1}}\rangle^{+}$ & $\alpha\vert 1\rangle_{B} + \beta\vert 0\rangle_{B}$ & $X$ gate\\
\hline
$\vert \tilde{w_{1}}\rangle^{-}$ & $\alpha\vert 1\rangle_{B} - \beta\vert 0\rangle_{B}$ & $-iY$ gate\\
\hline
$\vert \tilde{w_{2}}\rangle^{+}$ & $\alpha\vert 0\rangle_{B} + \beta\vert 1\rangle_{B}$ & $I$ gate (i.e. nothing to do) \\
\hline
$\vert \tilde{w_{2}}\rangle^{-}$ & $-\alpha\vert 0\rangle_{B} + \beta \vert 1\rangle_{B}$ & $-Z$ gate\\
\hline
\end{tabular}
\end{center}
\end{table}
\noindent Thus we see that our constructed state of Star class is useful as a quantum channel for standard teleportation.
\subsection*{$W_{n}$ state:}
\noindent We now take the general expression of non-prototype $W$ state defined in eq.(\ref{genw1}) (of which $W_{1}$ is a special case), only we consider the phase settings to zero so that this non-prototype state takes the following form.
\begin{eqnarray}
\label{non-prototypew1}
\vert W_{n}\rangle = \frac{1}{\sqrt{2+2n}}\Big[\vert 100\rangle + \sqrt{n}\vert 010\rangle + \sqrt{n+1}\vert 001\rangle\Big].
\end{eqnarray}
Its spin-flipped version, however, is
\begin{eqnarray}
\label{non-prototypew2}
\vert \tilde{W_{n}}\rangle = \frac{1}{\sqrt{2+2n}}\Big[\vert 011\rangle + \sqrt{n}\vert 101\rangle + \sqrt{n+1}\vert 110\rangle\Big].
\end{eqnarray}
Consequently, we can construct a state by taking equal linear superposition of states (\ref{non-prototypew1}) and (\ref{non-prototypew2}). The state is defined as
\begin{eqnarray}
\label{non-prototypew3}
\vert W_{n}\tilde{W_{n}}\rangle = \frac{1}{\sqrt{2}}\Big[\vert W_{n}\rangle + \vert \tilde{W_{n}}\rangle\Big].
\end{eqnarray}
For all the above states, we assume that first two qubits are taken by Alice and third qubit is taken by Bob while Alice wants to send unknown qubit (\ref{singlequbit}) to Bob. \\\\
Now for the state (\ref{non-prototypew1}) to be used as quantum teleportation channel the following basis will be considered.
\begin{eqnarray}
\label{basisnonpro1}
\vert w_{x}\rangle^{\pm} &=& \frac{1}{\sqrt{2+2n}}\Big(\vert 010\rangle + \sqrt{n}\vert 001\rangle \pm  \sqrt{n+1}\vert 100\rangle\Big)\nonumber\\
\vert w_{y}\rangle^{\pm} &=& \frac{1}{\sqrt{2+2n}}\Big(\vert 110\rangle + \sqrt{n}\vert 101\rangle \pm  \sqrt{n+1}\vert 000\rangle\Big).
\end{eqnarray}
With this basis and proceeding as before the following are the settings (as shown in table $VI$) for standard teleportation with state (\ref{non-prototypew1}).
\begin{table}[h!]
\begin{center}
\caption{Summary of perfect teleportation using $W_{n}$ state}
\label{table6}
\begin{tabular}{|c|c|c|}
\hline
Alice's measurement outcome & What Bob gets & What Bob needs to apply\\
\hline
$\vert w_{x}\rangle^{+}$ & $\alpha\vert 0\rangle_{B} + \beta\vert 1\rangle_{B}$ & $I$ gate (i.e. nothing to do)\\
\hline
$\vert w_{x}\rangle^{-}$ & $\alpha\vert 0\rangle_{B} - \beta\vert 1\rangle_{B}$ & $Z$ gate\\
\hline
$\vert w_{y}\rangle^{+}$ & $\alpha\vert 1\rangle_{B} + \beta\vert 0\rangle_{B}$ & $X$ gate \\
\hline
$\vert w_{y}\rangle^{-}$ & $-\alpha\vert 1\rangle_{B} + \beta \vert 0\rangle_{B}$ & $iY$ gate\\
\hline
\end{tabular}
\end{center}
\end{table}\\\\
Similarly for the state defined in (\ref{non-prototypew2}) to be used as quantum teleportation channel, the following basis will be taken into consideration.
\begin{eqnarray}
\label{basisnonpro2}
\vert \tilde{w_{x}}\rangle^{\pm} &=& \frac{1}{\sqrt{2+2n}}\Big(\vert 001\rangle + \sqrt{n}\vert 010\rangle \pm  \sqrt{n+1}\vert 111\rangle\Big)\nonumber\\
\vert \tilde{w_{y}}\rangle^{\pm} &=& \frac{1}{\sqrt{2+2n}}\Big(\vert 101\rangle + \sqrt{n}\vert 110\rangle \pm  \sqrt{n+1}\vert 011\rangle\Big).
\end{eqnarray}
The table $VII$ shows the measurements of Alice and subsequent application of appropriate unitary gates by Bob for the successful completion of the standard teleportation using the given state (\ref{non-prototypew2}) as channel.
\begin{table}[h!]
\begin{center}
\caption{Summary of perfect teleportation using $\tilde{W_{n}}$ state}
\label{table7}
\begin{tabular}{|c|c|c|}
\hline
Alice's measurement outcome & What Bob gets & What Bob needs to apply\\
\hline
$\vert \tilde{w_{x}}\rangle^{+}$ & $\alpha\vert 1\rangle_{B} + \beta\vert 0\rangle_{B}$ & $X$ gate\\
\hline
$\vert \tilde{w_{x}}\rangle^{-}$ & $\alpha\vert 1\rangle_{B} - \beta\vert 0\rangle_{B}$ & $-iY$ gate\\
\hline
$\vert \tilde{w_{y}}\rangle^{+}$ & $\alpha\vert 0\rangle_{B} + \beta\vert 1\rangle_{B}$ & $I$ gate (i.e. nothing to do)\\
\hline
$\vert \tilde{w_{y}}\rangle^{-}$ & $-\alpha\vert 0\rangle_{B} + \beta \vert 1\rangle_{B}$ & $-Z$ gate\\
\hline
\end{tabular}
\end{center}
\end{table}\\\\
Using the basis defined in eqs.(\ref{basisnonpro1}) and (\ref{basisnonpro2}), as before, one can use the state defined in (\ref{non-prototypew3}) for perfect/standard teleportation and the table $VIII$ gives the settings which the sender (Alice) and receiver (Bob) must follow.
\begin{table}[h!]
\begin{center}
\caption{Summary of perfect teleportation using $W_{n}\tilde{W_{n}}$ state}
\label{table1}
\begin{tabular}{|c|c|c|}
\hline
Alice's measurement outcome & What Bob gets & What Bob needs to apply\\
\hline
$\vert w_{x}\rangle^{+}$ & $\alpha\vert 0\rangle_{B} + \beta\vert 1\rangle_{B}$ & $I$ gate (i.e. nothing to do)\\
\hline
$\vert w_{x}\rangle^{-}$ & $\alpha\vert 0\rangle_{B} - \beta\vert 1\rangle_{B}$ & $Z$ gate\\
\hline
$\vert w_{y}\rangle^{+}$ & $\alpha\vert 1\rangle_{B} + \beta\vert 0\rangle_{B}$ & $X$ gate \\
\hline
$\vert w_{y}\rangle^{-}$ & $-\alpha\vert 1\rangle_{B} + \beta \vert 0\rangle_{B}$ & $iY$ gate\\
\hline
$\vert \tilde{w_{x}}\rangle^{+}$ & $\alpha\vert 1\rangle_{B} + \beta\vert 0\rangle_{B}$ & $X$ gate\\
\hline
$\vert \tilde{w_{x}}\rangle^{-}$ & $\alpha\vert 1\rangle_{B} - \beta\vert 0\rangle_{B}$ & $-iY$ gate\\
\hline
$\vert \tilde{w_{y}}\rangle^{+}$ & $\alpha\vert 0\rangle_{B} + \beta\vert 1\rangle_{B}$ & $I$ gate (i.e. nothing to do) \\
\hline
$\vert \tilde{w_{y}}\rangle^{-}$ & $-\alpha\vert 0\rangle_{B} + \beta \vert 1\rangle_{B}$ & $-Z$ gate\\
\hline
\end{tabular}
\end{center}
\end{table}\\\\
The above analysis shows that for all $n\geq 1$ (with phase settings considered zero), the state $W_{n}$, $\tilde{W_{n}}$ and $W_{n}\tilde{W_{n}}$ are useful for perfect teleportation.\\\\
Agarwal and Pati already shown that non-prototype $W$ state was useful as teleportation channel. From the above analysis we have shown that our constructed state $S_{w\tilde{w}}$, which is a Star type state, is also useful as teleportation channel.
\section{Analysis of non-prototype $W$ and $Star$ class of states:}
\noindent We now analyze the characteristics of the states, non -prototype $W_{1}$ state and $Star^1$ along with that of our constructed $S_{w\tilde{w}}$ state. We know that Star states have coherence and correlations distributed at all possible levels but in asymmetric way. The states have central qubit i.e. losing that qubit, the state becomes separable and a pair of peripheral qubits, loss of which do not affect the entanglement of the states. We summarize below the the coherence and concurrence of the states, $W_{1}$, $Star^1$ and $ S_{w\tilde{w}}$. For the calculations of coherence we have used $l_1-$ norm and for  bipartite correlations we used concurrence ($C$) and negativity ($N$). 
\subsection*{Coherence:}
\noindent A  widely used quantum coherence quantifier is the $\l_{1}$-norm measure and in our work we use the $C_{\ell_{1}}$ to represent this.   Since the 
quantum coherence is a basis dependent quantity, we fix the reference basis  $\ket{i}$ to a computational basis for a given quantum state.    The $l_{1}-$ norm of coherence is  defined 
as\textcolor{blue}{\cite{t2014}}.
\begin{eqnarray}
\label{l1norm}
C_{\ell_{1}}(\rho) = \sum_{i,j; i\ne j} \vert \rho_{ij}\vert, 
\end{eqnarray}
where $\rho_{ij} = \langle i\vert \rho\vert j\rangle$ is the matrix element corresponding to the $i^{th}$ row and $j^{th}$ column. Let us denote the density matrix corresponding to the state $Star^1$ by $\rho_{Star^1}$. We now use the eq.(\ref{l1norm}) to calculate the $1$ qubit ($\rho_{Star^1}^A$ etc), $2$ qubits ($\rho_{Star^1}^{AB}$ etc)   and $3$ qubits ($\rho_{Star^1}^{ABC}$)  coherence of the state $Star^1$. Thus we have
\begin{eqnarray}
\label{li1normstar1}
C_{\ell_{1}}(\rho_{Star^1}^A) &=& C_{\ell_{1}}(\rho_{Star^1}^B) = C_{\ell_{1}}(\rho_{Star^1}^C) = \frac{1}{2},\nonumber\\
C_{\ell_{1}}(\rho_{Star^1}^{AB}) &=& 1,\nonumber\\
C_{\ell_{1}}(\rho_{Star^1}^{AC}) &=& C_{\ell_{1}}(\rho_{Star^1}^{BC}) = \frac{3}{2},\nonumber\\
C_{\ell_{1}}(\rho_{Star^1}^{ABC}) &=& 3.
\end{eqnarray}
For the state $S_{w\tilde{w}}$, however, using eq.(\ref{l1norm}) we get
\begin{eqnarray}
\label{li1normstar2}
C_{\ell_{1}}(\rho_{S_{w\tilde{w}}}^A) &=& C_{\ell_{1}}(\rho_{S_{w\tilde{w}}}^B)= \frac{1}{\sqrt{2}},\nonumber\\
C_{\ell_{1}}(\rho_{S_{w\tilde{w}}}^C) &=& \frac{1}{2},\nonumber\\
C_{\ell_{1}}(\rho_{S_{w\tilde{w}}}^{AB})&=& C_{\ell_{1}}(\rho_{S_{w\tilde{w}}}^{AC})=C_{\ell_{1}}(\rho_{S_{w\tilde{w}}}^{BC}) =\sqrt{2} + \frac{1}{2},\nonumber\\
C_{\ell_{1}}(\rho_{S_{w\tilde{w}}}^{ABC})&=& 2+2\sqrt{2}.
\end{eqnarray}
This shows that the states $Star^1$ and $ S_{w\tilde{w}}$ have coherences distributed at all possible levels.\\\\
We find the coherence of state (\ref{genw2}) representing non-prototype $W$ state at all possible levels  (the density operator representation of this state is denoted by $\rho_{W_{1}}^{ABC}$). 
\begin{eqnarray}
\label{li1normstar1}
C_{\ell_{1}}(\rho_{W_{1}}^{A}) &=& C_{\ell_{1}}(\rho_{W_{1}}^{B}) = C_{\ell_{1}}(\rho_{W_{1}}^{C}) = 0,\nonumber\\
C_{\ell_{1}}(\rho_{W_{1}}^{AB}) &=& \frac{3}{4},\nonumber\\
C_{\ell_{1}}(\rho_{W_{1}}^{AC}) &=& \frac{1}{4} + \frac{1}{\sqrt{2}},\nonumber\\
C_{\ell_{1}}(\rho_{W_{1}}^{BC}) &=& \frac{1}{\sqrt{2}},\nonumber\\
C_{\ell_{1}}(\rho_{W_{1}}^{ABC}) &=& \frac{1}{2} + \sqrt{2}.
\end{eqnarray}
\subsection*{Concurrence:} 
\noindent Entanglement is an important quantum resource which arises due to nonlocal correlations between quantum sytems. To quantify entanglement of the bipartite versions of the aforementioned tripartite states we use concurrence, which for a quantum state $\rho$ \textcolor{blue}{\cite{wootters1998}} is 
\begin{eqnarray}
\label{concurrence}
C(\rho) = \max \lbrace 0, \sqrt{\lambda_{1}}-\sqrt{\lambda_{2}}-\sqrt{\lambda_{3}}-\sqrt{\lambda_{4}}\rbrace,  
\end{eqnarray}
where $\lambda_{1}\ge\lambda_{2}\ge\lambda_{3}\ge\lambda_{4}$ are the eigenvalues of the matrix $\rho \tilde{\rho}$.  The spin-flipped density matrix $\tilde{\rho}$ is
\begin{eqnarray}
\label{spin-flipped}
\tilde{\rho} = (\sigma_{y}\otimes \sigma_{y})\rho^{*}(\sigma_{y}\otimes \sigma_{y}),
\end{eqnarray}
where $\sigma_{y}$ is the Pauli spin matrix in the $y$-basis and $\tilde{\rho}$ is in the same basis as $\rho$, where  $\rho^{*}$ is the complex conjugate of the density matrix $\rho$.   Using (\ref{concurrence}), we have
\begin{eqnarray}
\label{constar1}
C(\rho_{Star^1}^{AB})&=& 0,\nonumber\\ 
C(\rho_{Star^1}^{AC}) &=& C(\rho_{Star^1}^{BC}) = \frac{1}{2},\nonumber\\
C(\rho_{S_{w\tilde{w}}}^{AB})&=& 0,\nonumber\\ 
C(\rho_{S_{w\tilde{w}}}^{AC}) &=& C(S_{w\tilde{w}}^{BC}) = \frac{1}{2}.
\end{eqnarray}
Also for non-prototype $W$ state (\ref{genw2}) we have
\begin{eqnarray}
\label{concw1}
C(\rho_{W_{1}}^{AB})&=& \frac{1}{2}\nonumber\\ 
C(\rho_{W_{1}}^{BC}) &=& C(\rho_{W_{1}}^{AC}) =\frac{1}{\sqrt{2}}.
\end{eqnarray}
\subsection*{Negativity:}
\noindent Another way to quantify entanglement is to find the negativity of the given state. Though not widely used as concurrence, for pure states negativity and concurrence give similar quantification. Negativity is a metric that quantifies the level of entanglement between two qubits, providing a numerical value for their quantum correlation. For a density matrix $\rho$ associated with a state, the negativity is denoted by $N(\rho)$ and the expression for evaluation of negativity is \textcolor{blue}{\cite{vidal2002}}
\begin{eqnarray}
\label{negativity}
N(\rho) &=& \frac{1}{2}\Big(\|\rho^{T_{A}}\|_{1} - 1\Big) = \vert \sum_{i} \lambda_{i}\vert = \frac{1}{2}\sum_{j}\Big[\Big(\vert \lambda_{j}\vert - \lambda_{j}\Big)\Big].
\end{eqnarray}
where, $\rho^{T_{A}}$ is the partial transpose of the composite system $\rho^{AB}$, (taken with respect to first party $A$). Here in above expression subscript $i$ runs over the subset of negative eigenvalues of $\rho^{T_{A}}$ whereas subscript $j$ runs over all the eigenvalues of $\rho^{T_{A}}$.\\\\
Using (\ref{negativity}), the negativities for the states defined in (\ref{star1}) and (\ref{starwwbar}) are given as
\begin{eqnarray}
\label{negstar}
N(\rho_{Star^1}^{AB}) &=& 0\nonumber\\
N(\rho_{Star^1}^{BC}) &=& N(\rho_{Star^1}^{AC}) = 0.232\nonumber\\
N(\rho_{S_{w\tilde{w}}}^{AB}) &=& 0\nonumber\\
N(\rho_{S_{w\tilde{w}}}^{BC}) &=& \rho_{S_{w\tilde{w}}}^{AC} = 0.232.
\end{eqnarray}
Also we have the negativities of two-qubits non-prototype $W$ state (which was defined in eq.(\ref{genw2})) and thereby are listed below.
\begin{eqnarray}
\label{negw1}
N(\rho_{W_{1}}^{AB}) &=& 0.103\nonumber\\
N(\rho_{W_{1}}^{BC}) &=& N(\rho_{W_{1}}^{AC}) = 0.25.\nonumber\\
\end{eqnarray}
We summarize the concurrence and negativity results of the aforementioned states in a tabular form below.
\begin{table}[h!]
\begin{center}
\caption{Summary of concurrence and negativity of tripartite states}
\label{table1}
\begin{tabular}{|c|c|c|c|}
\hline
State & Concurrence & Negativity\\
\hline
$\vert W_{1}\rangle$ & $C(\rho_{W_{1}}^{AB}) = 0.5$, $C(\rho_{W_{1}}^{BC}) =C(\rho_{W_{1}}^{AC})= 0.71$ &  $N(\rho_{W_{1}}^{AB}) = 0.103$,~~$N(\rho_{W_{1}}^{BC}) =N(\rho_{W_{1}}^{AC})= 0.25$\\
\hline
$\vert Star^1\rangle$ & $C(\rho_{Star^{1}}^{AB}) = 0$,~~$C(\rho_{Star^{1}}^{BC}) =C(\rho_{Star^{1}}^{AC})= 0.5$ & $N(\rho_{Star^{1}}^{AB}) = 0$,~~$N(\rho_{Star^{1}}^{BC}) =N(\rho_{Star^{1}}^{AC})= 0.232$\\
\hline
$\vert S_{w\tilde{w}}\rangle$ & $C(\rho_{S_{w\tilde{w}}}^{AB}) = 0$,~~$C(\rho_{S_{w\tilde{w}}}^{BC}) =C(\rho_{S_{w\tilde{w}}}^{AC})= 0.5$ & $N(\rho_{S_{w\tilde{w}}}^{AB}) = 0$,~~$N(\rho_{S_{w\tilde{w}}}^{BC}) =N(\rho_{S_{w\tilde{w}}}^{AC})= 0.232$\\
\hline
\end{tabular}
\end{center}
\end{table}
It is clear from the table $IX$ that, our constructed state $S_{w\tilde{w}}$ belongs to Star class of states as we see that when central qubit $C$ is removed both bipartite concurrence and negativity vanish. 
\section{Groverian and Tangle of tripartite states:}
\noindent We have already shown above that the states $Star^1$ and $S_{w\tilde{w}}$ are successful channels for standard (perfect) teleportation. In this section we shall make a comparative study of tripartite states concerned in this research work with respect to two measures of multipartite entanglement, one of which is tangle that gives genuine tripartite entanglement and another is an operational measure known as Groverian measure of entanglement. We will define these measures first. 
\subsection*{Tangle:}
A tripartite (or $3$ qubit) system in a pure state is generally represented as
\begin{eqnarray}
\label{tangledef1}
\vert \Psi\rangle_{ABC} = \sum _{i,j,k = 0}^{1}a_{ijk}\vert ijk\rangle
\end{eqnarray}
Here the qubits are distributed among three parties $A$, $B$ and $C$ and the state is written in computational basis $\lbrace \vert ijk\rangle\rbrace$. The complex coefficients $a_{ijk}$ obey a normalization relation. A genuine entanglement of such a three qubit system, defined in (\ref{tangledef1}), is quantified by tangle, denoted by $\tau$. The tangle $\tau$ is defined as\textcolor{blue}{\cite{coffman2000}}
\begin{eqnarray}
\label{tangledef2}
\tau(\vert \Psi\rangle_{ABC}) &=& C^2_{A(BC)} - C^2_{AB} - C^2_{BC}\nonumber\\
&=& 4 \vert d_{1} + 4 d_{2} - 2d_{3} \vert,
\end{eqnarray}
where $C^2$ is concurrence squared and 
\begin{eqnarray}
\label{tangledef2}
d_{1} &=& a_{000}^2a_{111}^2 + a_{001}^2a_{110}^2+a_{010}^2a_{101}^2 + a_{100}^2a_{011}^2, \nonumber\\
d_{2} &=& a_{000}a_{110}a_{101}a_{011} +  a_{111}a_{001}a_{010}a_{100},\nonumber\\
d_{3} &=& a_{000}a_{111}a_{011}a_{100} +a_{000}a_{111}a_{101}a_{010} + a_{000}a_{111}a_{110}a_{001}\\
&+& a_{011}a_{100}a_{101}a_{010} + a_{011}a_{100}a_{110}a_{001} + a_{101}a_{010}a_{110}a_{001}.
\end{eqnarray}
\subsection*{Groverian entanglement:}
\noindent A well-known measure which was constructed by an operational method from Grover's search algorithm is called Groverian measure. The measure for a state $\psi$ is defined as\textcolor{blue}{\cite{shimoni2004}}
\begin{eqnarray}
\label{grover1}
G(\vert \psi\rangle) = \sqrt{1 - P_{max}},
\end{eqnarray}
where,
\begin{eqnarray}
\label{grover2}
P_{max} = \max _{\vert e_{1}\rangle,\cdots, \vert e_{n}\rangle} \vert \langle e_{1}\vert \otimes \cdots \otimes \langle e_{n}\vert \psi\rangle\vert^2.
\end{eqnarray}
Physically $P_{max}$ corresponds to the maximal probability of success in Grover's search algorithm when $\vert \psi\rangle$ is an $n-$ qubit initial state.
Eylee Jung \textit{et.al} conjectured that for perfect teleportation,  the Groverian entanglement of the entanglement resource is $\frac{1}{\sqrt{2}}$ i.e. $0.71$. They stated that $P_{max} = \frac{1}{2}$ is a necessary and sufficient condition for the perfect two-party teleportation. Below we tabulate the states that have been considered here as quantum teleportation channels. We calculate the tangle ($\tau$), $P_{max}$ and  Groverian entanglement ($G$) of the given states using eq.(\ref{tangledef1}), (\ref{grover1}) and (\ref{grover2}).\\\\
\begin{table}[h!]
\begin{center}
\caption{Summary of tangle and Groverian entanglement of tripartite states}
\label{table1}
\begin{tabular}{|c|c|c|c|}
\hline
State & $P_{max}$ & Groverian ($G$) & Tangle ($\tau$)\\
\hline
$\vert GHZ\rangle$ & $\frac{1}{2}$ & $\frac{1}{\sqrt{2}}$ & $1$\\
\hline
$\vert W_{1}\rangle$ & $\frac{1}{2}$ & $\frac{1}{\sqrt{2}}$ & $0$\\
\hline
$\vert \tilde{W_{1}}\rangle$ & $\frac{1}{2}$ & $\frac{1}{\sqrt{2}}$ & $0$\\
\hline
$\vert W_{n}\rangle$ & $\frac{1}{2}$ & $\frac{1}{\sqrt{2}}$ & $0$\\
\hline
$\vert \tilde{W_{n}}\rangle$ & $\frac{1}{2}$ & $\frac{1}{\sqrt{2}}$ & $0$\\
\hline
$\vert W_{n}\tilde{W_{n}}\rangle$ & $\frac{1}{4}$ & $\frac{\sqrt{3}}{2}$ & $\frac{1}{2(n+1)}$\\
\hline
$\vert S_{w\tilde{w}}\rangle$ & $\frac{1}{4}$ & $\frac{\sqrt{3}}{2}$ & $\frac{1}{4}$\\
\hline
$\vert Star^1\rangle$ & $\frac{1}{4}$ & $\frac{\sqrt{3}}{2}$ & $\frac{1}{4}$\\
\hline
\end{tabular}
\end{center}
\end{table}
From table $X$ below it is clear that, for the star-type states $Star^1$ and $S_{w\tilde{w}}$, $P_{max} \neq \frac{1}{2}$, rather less than this. So these two types of states are not $P_{max}= \frac{1}{2}$ states according to Jung conjecture. Yet the states can successfully be used as channels for standard (perfect) teleportation. Also we observe that the tangles of $W_{1}$ state and its spin-flipped version are $0$ though both the states are useful in successful perfect teleportation. This implies that genuine tripartite entanglement is not a necessary requirement for perfect teleportation.  The tangles of $Star^1$ and $S_{w\tilde{w}}$ states are however $\frac{1}{4}$. It is also clear from the table that the tangle $\tau$ of $W_{n}\tilde{W_{n}}$ is $\frac{1}{2(n+1)}$ while that of $W_{1}\tilde{W_{1}}$ is $\frac{1}{4}$ which is for $n=1$. Moreover, $G(W_{n}\tilde{W_{n}})$ is $\frac{\sqrt{3}}{2}$, which is independent of $n$.
\section{Conclusion:}
\noindent In this work, we have analyzed the efficacy of certain tripartite quantum pure states considered as shared channels between Alice and Bob. Although these states are $3$ qubit states, we have involved two parties Alice and Bob only while Alice holds two qubits and Bob holds one. First we have shown that with $GHZ$ state perfect teleportation is possible where Alice plays the role of sender holding an unknown qubit and Bob is the receiver. It is known that prototype $W$ state is not useful for standard teleportation. Hence we have taken non-prototype $W_{1}$ state as defined in eq.(\ref{genw2}), and used it as quantum teleportation channel using the protocol described by Agarwal and Pati \cite{pati2006}. Consequently we have studied the spin-flipped version of this non-prototype $W_{1}$ state i.e. $\tilde{W_{1}}$, described in eq.(\ref{genw3}) and have shown that the state is also useful for standard teleportation channel. The linear superposition of the $W_{1}$ and $\tilde{W_{1}}$ i.e. $S_{w\tilde{w}}$ of eq.(\ref{starwwbar}),  is then taken into consideration and its efficacy as quantum teleportation channel is verified. We also have shown that $Star$ type state defined in eq.(\ref{star1}) also serves well as quantum channel for standard teleportation. It is interesting to note down that our constructed state $S_{w\tilde{w}}$ belongs to a Star class of states. Then we have studied tangle and Groverian entanglement of these tripartite states. It is found that the states $W_{1}$ and $\tilde{W_{1}}$ have zero tangle and non-zero Groverian entanglement. But $Star^1$ and $S_{w\tilde{w}} $ have both non-zero tangle and Groverian entanglement. Groverian entanglements of $Star^1$ and $S_{w\tilde{w}}$ are greater than the tangles of the states. Also it is observed that genuine tripartite entanglement as measured by tangle is not significant for a state to be successful as quantum teleportation channel. We have already seen that $W_{1}$ and $\tilde{W_{1}}$ states are efficient quantum teleportation channel with tangles for each as zero. Our analysis also have revealed an important fact. As conjectured by Jung \textit{et.al}, a state when used as quantum teleportation channel, must have its Groverian entanglement as $\frac{1}{\sqrt{2}}$ or in other words the states are $P_{max}=\frac{1}{2}$ states. According to them this condition is necessary and sufficient. We have shown that this is not the case. Our constructed state $S_{w\tilde{w}}$ and $Star^1$ are not $P_{max}=\frac{1}{2}$ states but are useful for perfect two party teleportation. In future work we can study more generally the class of $Star$ states for successful quantum information processing tasks such as teleportation. Furthermore, one can make the analysis of superdense coding with the $Star$ type of states discussed in this work as well as with the general class of $Star$ type of states. Experimental realization of our constructed state $S_{w\tilde{w}}$ state of eq.(\ref{starwwbar}) can also be pursued in future works.
\vskip 0.5cm
{\bf Declaration of competing interest} The authors declare that they have no known competing financial interests or 
personal relationships that could have appeared to influence the work reported in this paper.
\vskip 0.5cm
{\bf Data availability statement} All data that support the findings of this study are included within the article. No supplementary file has been added.

\end{document}